\documentclass[aps,prl,twocolumn,groupedaddress,showpacs]{revtex4}
\usepackage{graphicx}
\begin{document}

%
%
%
%
\title{ Proof of principle for a high sensitvity search for the electric
dipole moment of the electron using the metastable 
$a(1) \left[ {^3 \Sigma ^+} \right]$ state of PbO}
%
%
%
%
\def\a1{$a(1)\left[^{3}\Sigma^{+}\right]$}
\def\X0{$X(0)\left[^{1}\Sigma^{+}\right]$}
\author{D. Kawall, F. Bay, S. Bickman, Y. Jiang, and D. DeMille}
%
%
%
%
\affiliation{Department of Physics, Yale University, P.O. Box 208120,
New Haven, CT 06520-8120}
%
%
%
%
\date{\today}
\begin{abstract}
The metastable \a1 state of PbO has been suggested as a suitable
system in which to search for the electric dipole moment (EDM) of
the electron.  We report here the development of experimental
techniques allowing high-sensitivity measurements of Zeeman and
Stark effects in this system, similar to those required for an EDM
search. We observe Zeeman quantum beats in fluorescence from a
vapor cell of PbO, with shot-noise limited extraction of the
quantum beat frequencies, high counting rates, and long coherence
times. We argue that improvement in sensitivity to the electron
EDM by at least two orders of magnitude appears possible using
these techniques.
\end{abstract}
%
%
%
%
%
%
%
\maketitle
%
%
A permanent electric dipole moment (EDM) of the electron $d_{e}$ would
violate parity and time-reversal symmetries \cite{khripandlam}. 
Many extensions of the standard model predict $d_{e}$ within 1-3 
orders of magnitude of the current limit \cite{regan1}; thus extending 
the sensitivity to $d_{e}$ by a few orders of magnitude may offer the 
exciting possibility of observing new physics beyond the 
standard model \cite{commins1}.
 
There is a dramatic advantage to using heavy polar molecules in such 
searches \cite{sandars1}. Strong hybridization of
the atomic orbitals in such molecules leads to enormous internal
electric fields along the internuclear axis $\hat{\bf n}$.  
The close spacing of levels of opposite parity allows full polarization of 
this axis along
external fields attainable in the laboratory, enhancing the linear
Stark effect produced by $d_{e}$ two or more orders of
magnitude beyond that expected in atoms \cite{misha1}. Still, this
theoretical advantage has yet to translate into a competitive
limit on $d_{e}$.

In this paper we describe several measurements which constitute a
proof of principle for a new type of search for $d_{e}$.  Our
experiment uses the metastable \a1 state of PbO. 
We have earlier proposed to use this state to extend the current bound 
$|d_{e}|\le 1.6\times 10^{-27}~e\cdot$cm \cite{regan1} 
by 2-4 orders of magnitude \cite{dave1}.
The high sensitivity of the proposed EDM experiment relies on
several properties of the $a(1)$ state of PbO, including its large
enhancement factor \cite{misha2, titov1}, 
and unique possibilities for the rejection of
systematic errors \cite{davecommins}. The key technical advance
reported here is the ability to create, manipulate, and measure
coherent electron spin superpositions in a high-density PbO 
vapor cell.  This leads to measurements of Zeeman and
Stark splittings, like those for an EDM search,
with narrow linewidths and at high signal-to-noise $S/N$.  
With this ability we measured coherence
times and quenching cross sections in the vapor cell;
demonstrated sensitive, shot-noise limited extraction of Zeeman
quantum beat frequencies; and measured Land\'e $g$-factors.
Finally, we have demonstrated the
ability to apply electric fields sufficiently strong to fully
polarize the $a(1)$ state.

A barrier to molecule-based searches for $d_{e}$ has been 
achieving sufficient numbers of suitable molecules -- count rates of 
$10^{4}$ Hz limited the sensitivity of a recent molecular beam experiment 
using YbF \cite{hinds1}, which was far from systematic limits.
To achieve much higher counting rates we developed 
a specialized oven to heat a novel vapor cell to
temperatures of  $700 ^\circ \rm{C}$, where PbO has
substantial vapor density, $n\approx 4\times 10^{13}~$cm$^{-3}$.
The cell contains about 80 cm$^{3}$ of PbO vapor of
natural isotopic abundance, and has gold foil electrodes for measurements
requiring an electric field. The cell and oven sit
in a vacuum chamber surrounded by 3 orthogonal Helmholtz coils. 

With this apparatus we can populate and detect fluorescence from the 
metastable state of interest as follows.
We populate the $a(1)(v'=5)$ state of $^{208}$PbO by laser
excitation from the ground state $X(0)(v''=1)$.  
A dye laser is pumped by the second harmonic of a Nd:YAG laser
at a rep. rate of 100 Hz, and delivers
10-20 mJ/pulse of light at wavelength $\lambda \approx 571$ nm
propagating in the $\hat{\bf y}$ direction to the vapor
cell. The pulses are $\approx 8$ ns long with a linewidth of
1 GHz, comparable to the Doppler width of the transition. The
light traverses the vacuum chamber and oven in a 5 cm diam. 
lightpipe to the cell, then exits
through another lightpipe. Fluorescence from the $a(1)(v'=5)
\rightarrow X(0)(v''=0)$ decay at $\lambda \approx 548
\rm{nm}$ is captured by a third lightpipe orthogonal to the
laser beam. The fluorescence passes through an infrared-blocking
filter and two interference filters ($550 \pm 5$ nm bandpass),
which block scattered laser light and most blackbody
radiation from the ovens, while passing the signal of interest to
a photomultiplier tube (PMT).

The EDM search and all measurements discussed here
use the levels of the $a(1)$ state with total
angular momentum $J$=1. 
This Hund's case (c) state can have projections $\Omega$ of 
electronic angular momentum ${\bf{J}}_{e}$ along or against the internuclear 
axis, $\Omega={\bf J}_{e}\cdot \hat{\bf n}=\pm1$. Even and odd combinations 
form nominally degenerate parity eigenstates. The degeneracy is broken by
the Coriolis coupling between electronic and rotational angular
momentum. This splits each level $J$ into two closely
spaced states of opposite parity, $e$ and $f$, with parity
$(-1)^{J}$ and $(-1)^{J+1}$ respectively \cite{brom_veseth} and separation  
$\Delta_{\Omega}(J) = qJ(J+1)$, where $q=5.6(1)$ MHz for 
$a(1)(\nu'=5)$ \cite{larry1}. 

In the presence of a static electric field,
a non-zero value of $d_{e}$ results in a linear Stark shift in the 
$M=\pm 1$ sublevels of these $J=1$ levels \cite{dave1,davecommins}. 
The current limit on $d_{e}$
corresponds to a 15-40 mHz shift \cite{misha2,titov1}.
To test the method proposed to detect this shift in the EDM search,
we studied Zeeman quantum beats in a static magnetic
field ${\bf B}=B\hat{\bf z}$
which shifts the $a(1)$ levels by
$g_{i}\mu_{B}BM/J(J+1)$. Here $\mu_{B}$ is the Bohr magneton, and 
$g_{i}=g_{e}~(g_{f})$ is the Land\'e
$g$-factor in the molecule-fixed frame of the $e~(f)$ member of the
$\Omega$-doublet.  

Pulses of $\hat{\bf x}$-polarized light were used to create a coherent
superposition of $M = \pm 1$ sublevels, equivalent to an alignment of
the angular momentum ${\bf J}$ along $\hat{\bf x}$.  The $B$-field
removes the $M=\pm 1$ degeneracy so the sublevels evolve at
different rates and acquire a phase difference, leading to a
precession of the alignment about $\hat{\bf z}$ \cite{corney}.
We detect 
this precession as $\Delta M=2$ quantum beats which 
modulate the exponential decay in the
unpolarized fluorescence intensity at twice the Larmor frequency,
$\nu_{b}=2g_{i}\mu_{B}B/hJ(J+1)$. 
\begin{figure}
\includegraphics[width=\columnwidth]{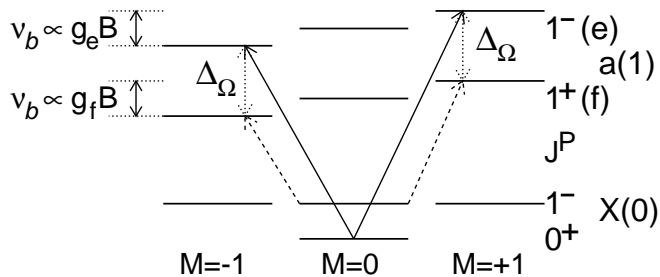} 
\caption{\label{fig:levels}
The energies of the low-lying X(0) and a(1) states in a magnetic 
field {\bf B} are shown. 
The coherent $R0$ ($Q1$) transition leading to quantum beats at 
$\nu_{b}\propto g_{e}B~(g_{f}B)$ is indicated by solid 
(dashed) arrows. 
The $\Omega$-doublet splitting $\Delta_{\Omega}$ 
(which is independent of {\bf B} and $M$) is shown greatly exaggerated. 
Transitions between the doublet levels can be driven by
an RF electric field at $\nu_{RF}=\Delta_{\Omega}$ (dotted arrows).
}
\end{figure}
Electric dipole selection rules allow us to selectively excite
either parity state of the $J=1$ $\Omega$-doublet.
This is accomplished by tuning the dye laser frequency to either
the $X(J=0^+) \rightarrow a(J=1^-)$ transition ($R0$ line), or the
$X(J=1^-) \rightarrow a(J=1^+)$ transition ($Q1$ line),
separated by $\approx 18 \rm~{GHz}$. The relevant levels and transitions 
are shown in Fig. \ref{fig:levels}.

\begin{figure}
\includegraphics[width=\columnwidth]{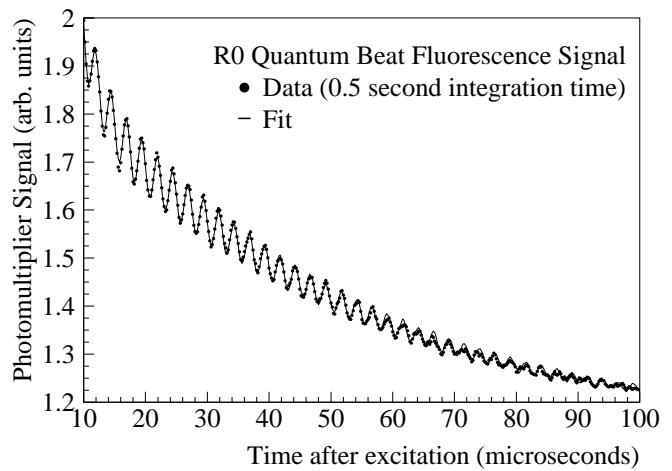}
\caption{\label{fig:beats} Zeeman quantum beat data representing 
0.5 seconds of integration are shown with a fit to Eq. (\ref{eqn:beatsig}).}
\end{figure}

The time-dependent quantum beat fluorescence signals, $I(t)$, were
fit to the form (see Fig. \ref{fig:beats}) :
\begin{eqnarray}
I(t)&=&\alpha S(t) \times \left[1 + c e^{-t/T_{b}}
\cos(2\pi\nu_b t+\phi)\right]+d+P(t)\nonumber\\
& &+\beta L(t). {\label{eqn:beatsig}}
\end{eqnarray}
The terms in this equation are as follows.  $P(t)$ is an
electronic transient associated with switching off the PMT gain
during the laser pulse, and $L(t)$ is the residual signal
associated with scattered laser light; both are recorded off-line.
$S(t)$ describes the fluorescence signal in the absence of quantum
beats; this was determined by applying an inhomogeneous magnetic
field $dB_{z}/dz\approx 0.2$ G/cm which caused rapid decoherence of
the beats.  $S(t)$ is approximately exponential in form, with
deviations due to wall quenching and time-dependent acceptance
changes resulting from diffusion of the excited molecules.  The remaining seven
parameters are adjusted to fit the quantum beat signals: scale factors 
$\alpha,\beta\approx 1$ for the signal
and scattered light, respectively; the beat contrast $c\approx
0.1$; a factor $T_{b}\approx 100~\mu$s, accounting for
shortening of the beat coherence time by collisions; the quantum
beat frequency $\nu_b \approx~125-400$ kHz (corresponding to
$B\approx 50-160$ mG); the beat phase $\phi$; and the 
background level due to blackbody radiation from the oven $d$; in practice
$d\approx S(0)$.  The fits start $10-15~\mu s$
after the laser pulse to lessen the impact of the transients on the fit.

We used data of this type to optimize the cell conditions
for the extraction of Zeeman beat frequencies.
Collisions with ground state PbO molecules can quench
molecules in the excited state and reduce the effective 
lifetime $T_{1}$ of the $a(1)$ state in the cell, and reduce the
quantum beat coherence time $T_{2}$. We expect
\begin{eqnarray}
\frac{1}{T_{1}}\simeq\frac{1}{\tau_{a}}+
\frac{1}{\tau_{\mathrm{cell}}}+\sigma_{1}nv,~\mathrm{and}~
\frac{1}{T_{2}}\simeq\frac{1}{T_{1}}+\sigma_{2}nv
\end{eqnarray}
where $\tau_{a}\sim 82(2)~\mu s$ is the natural
lifetime \cite{dave1}, $\tau_{\mathrm{cell}}$ is an approximate
time constant for quenching on the cell walls, $\sigma_{1}$ is the
state quenching cross-section, $v$
is the average relative velocity ($v \approx~400$ m/s) and
$\sigma_{2}$ is the cross-section for beat quenching but state
preserving collisions ($\sigma_{2}nv\simeq 1/T_{b}$ in
Eq. (\ref{eqn:beatsig})). The optimal $S/N$ in beats is obtained when the
density is adjusted so the
collisional decoherence rate is comparable to the decay rate of
the excited state, $(\sigma_{1}+\sigma_{2})nv\simeq
1/\tau_{a}+1/\tau_\mathrm{cell}$. We extracted $\sigma_{1}$ and $\sigma_{2}$
by approximating $S(t)$ with  
$\exp(-t/T_{1})$ in the fits to Eq. ({\ref{eqn:beatsig}}), 
then observing the changes in $T_{1}$ and $T_{2}$
as the number density in the cell was varied by more
than an order of magnitude (by changing the cell temperature). 
We find $\sigma_{1} \approx 0.4\times 10^{-14}$
cm$^{2}$ and $\sigma_{1}+\sigma_{2} \approx 1.8\times 10^{-14}$
cm$^{2}$, with a factor of two uncertainty,
reflecting the range of results obtained with different cells 
and from systematic uncertainties in the fitting.
The results imply an optimal cell density $n
\approx 3 \times 10^{13}$ cm$^{-3}$ (corresponding to a cell
temperature of 690$^{\circ}$C), enabling the experiment
to run near its originally proposed sensitivity \cite{dave1}.

We ran at this optimal condition to determine the noise in our
extraction of Zeeman beat frequencies.We typically achieved a
sensitivity $\delta \nu_b \approx
50~\mathrm{Hz}/\sqrt{\mathrm{Hz}}$ for integration periods $T<1$s.
Drift in the ambient magnetic field 
between laser shots dominated the uncertainty over
longer intervals. This short-term sensitivity matches the
shot-noise limited uncertainty in frequency
expected from :
\begin{eqnarray}
\delta\nu_b \approx \frac{\kappa\sqrt{2}}{2\pi c T_{2}\sqrt{\dot{N}}}\approx
\frac{50~\mathrm{Hz}}{\sqrt{\mathrm{Hz}}}
\end{eqnarray}
where $\sqrt{2}$ comes from fitting the phase;
$\kappa$ is an excess noise factor depending on the ratio $S(0)/d$; 
$\dot{N} \gtrsim 1\times 10^7/s$ is the average detected
fluorescence count rate, and $T_{2}\approx 50~\mu\rm{s}$. 
We find $\kappa\approx 3$ through modeling. 
We note that the measured value of $\dot{N}$ is close to 
expectations, based on the estimated efficiencies for our current
setup.  The excitation efficiency is $\varepsilon_{e}\approx 0.03\%$, 
based on the
available laser power, rep. rate, and estimated $X(0)-a(1)$
excitation cross-section \cite{dave1,davecommins}. The
detection efficiency is $\varepsilon_{d}\approx 0.003\%$, based on the PMT 
quantum efficiency and simulated geometric collection efficiency.

We used this apparatus to make precise measurements of $g_e$ and
$g_f$ for the $J=1$ levels of the $a(1)$ state. These quantities
are of interest for two reasons. The deviation of $\bar{g} = (g_e
+ g_f)/2$ from $2$ is a measure of spin-orbit mixing in $a(1)$,
and enters the semi-empirical estimate of the EDM enhancement
factor \cite{misha2}. The difference $\delta g = g_{e}-g_{f}$ is
important for determining the level at which some systematics in
the EDM measurement can be rejected by using the doublet levels as
internal co-magnetometers \cite{davecommins}.

We extracted $g$-factors from the 
changes observed in $\nu_b$ corresponding to controlled changes in $B$-field 
magnitude. These were determined before and after data-taking 
in the volume occupied by the cell with a calibrated 3-axis magnetometer. 
We found $g_e =~1.857(6)$, in agreement with our earlier, less
precise measurement \cite{larry1}. Our precision was limited primarily
by uncertainty in the magnetic
field calibration, with smaller uncertainties due to magnetic
field noise and uncertainties in the fit function. By driving the $Q1$ 
transition and comparing changes in beat frequencies with the $R0$ results 
for the same changes in $B$, the difference
in $g$-factors was determined : 
$\delta g =-25(13)\times 10^{-4}$. Here the uncertainty is
largely immune to the calibration, and comes from equal
parts statistical uncertainty and magnetic field noise.

We obtained higher accuracy in $\delta g$ using a more
sophisticated method.  As before, we drove the $R0$ transition
with $\hat{\bf x}$ polarized light, creating alignment of the
$J=1^- (e)$ level, and measured $\nu_b \propto g_e$.  Periodically, we also
created an alignment in the $J=1^+ (f)$ level, so $\nu_b \propto g_f$, 
in the following manner.  Three microseconds
after the laser pulse, we applied a pulse of RF
electric field at frequency $\nu_{RF}=\Delta_{\Omega}/h$, to
resonantly drive the $\Delta M=0$ electric dipole transition between 
the $e$ and $f$ levels
(see Fig. \ref{fig:levels}). The time duration 
$T_{RF}\approx 5~\mu$s and amplitude $E_{RF}\approx 0.12~$V/cm
of the RF pulse were adjusted to create a $\pi$-pulse.
This method allows us to switch more
rapidly between $e$ and $f$ levels than was possible by tuning the
laser between the $R0$ and $Q1$ transitions, and thus eliminates
much of the noise due to drifts in ambient $B$-field during the
measurement.  Measuring the change in beat frequency between data
with or without the RF pulse yields $\delta g =-31(9)
\times 10^{-4}$. The error is from uncertainty in the
efficiency of the $\pi$-pulse transfer, and from the statistical
uncertainty in the beat frequency differences. The combined result
is $\delta g =-30(8) \times 10^{-4}$, in coarse agreement with predictions. 

\begin{figure}
\includegraphics[width=\columnwidth]{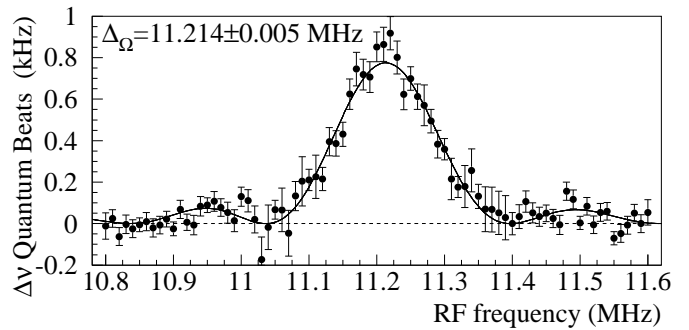}
\caption{\label{fig:omega}
The change in quantum beat frequency $\nu_{b}$ is shown as a function
of the frequency $\nu_{RF}$ of an RF $\pi$-pulse. A $\pi-$pulse at the 
frequency of the $\Omega$-doublet splitting 
transfers population from the $e$ to $f$ levels and changes the
beat frequency from $\nu_{b}\propto g_{e}B$ to $\nu_{b}\propto g_{f}B$.
} 
\end{figure}

This difference in $g$-factors was the basis for an 
RF-spectroscopic study of the $\Omega$-doublet splitting, by
monitoring the change in $\nu_b$ as a function of $\nu_{RF}$,
as shown in Fig. \ref{fig:omega}. 
We found $\Delta_{\Omega} = 11.214(5)
\rm{MHz}$, where the uncertainty is dominated by possible systematic
effects due to off-resonant AC Stark shifts (Bloch-Siegert shift).
This is consistent with the best previous measurement
\cite{larry1}, but 40 times more precise. We obtain similar results
by monitoring the change in beat contrast when 
driving the $e\rightarrow f$ transition. This contrast difference is
associated with different angular distributions of fluorescence for the
$e\rightarrow X(J''=0,2)$ and $f\rightarrow X(J''=1)$ transitions.

Next we turn to results obtained with a static electric field
${\bf E}=E\hat{\bf z}$. Such a field mixes the $e,f$ sublevels with the 
same value of $M$, leading to states of mixed parity separated in
energy by $\Delta(E)=\sqrt{\Delta_{\Omega}^2+(\mu_{J,M}E)^2}$.
Here $\mu_{J,M}$ is the matrix element of the electric dipole
operator $\hat{\mu}$: $\mu_{J,M} = \left\langle {e,J,M| \hat{\mu}
|f,J,M} \right\rangle = \mu_{a}M/J(J+1)$, and $\mu_{a}$=1.64(3)
MHz/(V/cm) is the molecule-fixed dipole moment in the $a(1)$ state
\cite{larry1}. 

We used the PbO molecules to sense weak applied $E$-fields ($E~<$2 V/cm),
by measuring the Stark shift in the
$\Omega$-doublet splitting.  As before, we populated the $J=1^{-}$
($e$) $M=\pm 1$ levels at $E=0$ by driving the $R0$ transition
with $\hat{\bf{x}}$ polarized light. Several microseconds after the laser
pulse, $E$ was ramped up adiabatically over 2.5 $\mu$s
($1/E\cdot dE/dt < \Delta_{\Omega}/h$), and an RF $\pi$-pulse
($\approx~5~\mu$s in duration) was applied to the electrodes at
$\nu_{RF}=\Delta_{\Omega}/h$ to drive the $E1~\Delta M=0$
transition to the other state of mixed parity. Then $E$ was ramped
down adiabatically and the quantum beat frequency was determined.
For $\nu_{RF}$ far from resonance, no population is transferred to
the lower level, so the beats are characterized by $g_{e}$,
whereas resonant excitation yields beats characterized by $g_{f}$.
The observed dependence of the splitting with electric field gives
experimental evidence of an electric field in the cell in qualitative 
agreement with predictions (limited by our knowledge of the
relation between applied voltage and $E$-field in the cell). 

This method for determining Stark shifts could not be extended to
larger $E-$fields 
due to technical limitations.
However, we observed clear signatures of large
$E$-fields in the cell.  For these observations,
$E$ was kept on during the laser excitation, which used
$\hat{\bf{x}}$-polarized light tuned to the $R0$ transition.  We
observed a minimum in beat lifetime $T_{2}$ when the Stark shift
approximately matches the Zeeman
shift, bringing the $M=-1$ and $M=0$ levels into degeneracy. Then
Majorana spin-flips depopulate the $M=-1$ level and
minimize the beat lifetime. 
Upon increasing $E$ the degeneracy is broken;
the beat lifetime recovers and is limited finally by field inhomogeneities. 
Large $E$-fields $E\ge h\Delta_{\Omega}/\mu_{a}$
should also result in a beat contrast equal to 
the average of $e$ and $f$ level contrasts, 
corresponding to a rapidly oscillating coherent superposition of these states. 
We observed contrasts approaching $80\%$ of this value. We attribute this
to difficulties distinguishing changes in contrast from lifetime.

Maximum sensitivity to an EDM requires applying a field
large enough to reach electrical polarization 
\hbox{$P=<{\bf J}_{e}\cdot\hat{\bf n}>\approx \pm 1$}, corresponding
to complete mixing of the $\Omega$-doublet states to form eigenstates
of definite $\Omega$ \cite{misha2}. Here $P$ depends on $E$ as
$P = 2\alpha\beta/(\alpha^2+\beta^{2})$ where $\alpha=\mu_{J,M}E$, and 
$\beta=\Delta_{\Omega}/2+\sqrt{(\Delta_{\Omega}/2)^2+(\mu_{J,M}E)^{2}}$. 
We have successfully applied static electric fields 
$E > 50~\rm{V/cm}$ to our cell, corresponding to $P\sim 99\%$,
with no signs of failure.  However, we observed a substantial 
leakage current in the cell 
$I_L \approx 1~\mu\rm{A}$ at $E \approx 50~\rm{V/cm}$.
In the worst case this current can produce a field $B_{L}\approx 10^{-7}$ G 
causing beat frequency shifts $\Delta\nu_{b}\approx \bar{g}\mu_{B}B_{L}/h$.
The shift changes sign upon reversal of $E$, and mimics an
EDM. We expect to reduce $I_{L}$ considerably with improved cell design and 
handling procedures.

A powerful technique for separating the EDM from such systematics 
comes through comparisons of $\nu_{b}$ observed in the two doublet 
levels, without the need to reverse fields \cite{davecommins}.  
Fully mixed upper and lower doublet levels 
have opposite orientations of the internal electric field and exhibit 
opposite shifts due to an EDM. Systematic shifts retain the same sign
and magnitude to the extent that the effective $g$-factors of the mixed
doublet levels are the same.
Thus in the limit of nearly complete mixing $\mu_{a}E\gg h\Delta_{\Omega}$, 
leakage currents change $\nu_{B}$ between doublet levels by only 
$\Delta\nu_{b}\approx \delta g\mu_{B}B_{L}[\Delta_{\Omega}/(h\mu_{a}E)]$, 
suppressing the systematic shift due to $B_{L}$ by $\gg 10^{3}$, and 
allowing a systematic limit $\lesssim~10^{-29}~e\cdot$cm.

In conclusion, we have demonstrated the feasibility of a new
approach to measuring the electric dipole moment of the electron
in the metastable \a1 state of PbO. 
With our present configuration, it is feasible to achieve EDM
sensitivity comparable to the current limit \cite{regan1}.
However, the sensitivity can be improved dramatically with
straightforward modifications. In particular: by using two
photodiodes with high quantum efficiency instead of a single PMT,
exciting from the $X(0)(\nu''=0)$ level, using broad band
interference filters to capture the fluorescence into two
vibrational levels simultaneously, using isotopically enriched
$^{208}$PbO, and other improvements, we expect to increase
the count rate by more than three orders of magnitude, and the contrast
by more than a factor of two.
These changes should result in a statistical sensitivity approaching
100 mHz/$\sqrt{\mathrm{Hz}}$. From the theoretical estimates of the
enhancement factor \cite{titov1,misha2}, a statistical limit at the level of
$|d_{e}|\sim 10^{-29}~e\cdot$cm will be achieved in less than a
month of integration.  The necessary modifications are now
underway.

\begin{acknowledgments}
We are grateful for the support of NSF Grant PHY9987846, a NIST Precision
Measurement Grant, Research Corporation, the David and Lucile Packard
Foundation and the Alfred P. Sloan Foundation.
\end{acknowledgments}


\begin{thebibliography}{99}
%
%
\bibitem{khripandlam}
I.B. Khriplovich and S.K. Lamoreaux, {\it{$CP$ Violation Without Strangeness}}
(Springer, New York 1997).
\bibitem{regan1}
B.C. Regan {\it{et al.}}, Phys. Rev. Lett. {\bf{88}}, 071805 (2002).
\bibitem{commins1}
Predictions for $d_{e}$ are reviewed in E.D. Commins,
Adv. At. Mol. Opt. Phys., {\bf{40}}, 1 (1999).
\bibitem{sandars1}
P.G.H. Sandars, Phys. Rev. Lett. {\bf{19}}, 1396 (1967).
\bibitem{misha1}
O.P. Sushkov and V.V. Flambaum, Zh. Eksp. Teor. Fiz. {\bf{75}}, 1208 (1978)
[Sov. Phys. JETP {\bf{48}}, 608 (1978)];
M.G. Kozlov and L.N. Labzowsky, J. Phys. B. {\bf{28}}, 1933 (1995).
\bibitem{dave1}
D. DeMille {\it{et al.}}, Phys. Rev. A {\bf{61}}, 052507 (2000).
\bibitem{misha2}
M.G. Kozlov and D. DeMille, Phys. Rev. Lett. {\bf{89}}, 133001 (2002).
\bibitem{titov1}
T.A. Isaev {\it{et al.}}, physics/0306071; A. Titov (private communication).
\bibitem{davecommins}
D. DeMille {\it{et al.}}, in {\it{Art and Symmetry in Experimental
Physics}}, ed. by D. Budker {\it {et al.}} (AIP, New York 2001).
\bibitem{hinds1}
J.J. Hudson {\it{et al.}},
Phys. Rev. Lett. {\bf{89}}, 023003 (2002).
\bibitem{brom_veseth}
J.M. Brom and W.H. Beattie, J. Mol. Spectrosc. {\bf{81}}, 445 (1990);
L. Veseth, J. Phys. B. {\bf{6}}, 1473 (1973).
\bibitem{larry1}
L.R. Hunter {\it{et al.}}, Phys. Rev. A {\bf{65}}, 030501(R) (2002).
\bibitem{corney}
A. Corney, {\it{Atomic and Laser Spectroscopy}} (Clarendon Press, Oxford 1977).
%
%
\end{thebibliography}
%
%

%
%
\end{document}